\documentclass[12pt]{article}
\usepackage{amssymb,amsmath}
\usepackage{graphicx}
\usepackage{epstopdf}
\usepackage[affil-it]{authblk}
\usepackage[top=0.75in, bottom=0.75in, left=0.75in, right=0.75in, dvips]{geometry}
\usepackage{caption}
\usepackage{color}
\usepackage{ulem}

\begin{document}
\textwidth 10.0in
\textheight 9.0in
\topmargin -0.60in
\title{Renormalization Group Summation of Laplace QCD Sum Rules for Scalar Gluon Currents}
\author{Farrukh Chishtie \thanks{fchishti@uwo.ca}}
\affil{Department of Space Science, Inst. of Space Technology, Islamabad 44000, Pakistan}
\author {T.G. Steele \thanks{Tom.Steele@usask.ca}}
\affil {Department of Physics and Engineering Physics, University of Saskatchewan, Saskatoon S7N 5E2, Canada }
\author{D.G.C. McKeon \thanks{dgmckeo2@uwo.ca}}
\affil {Department of Applied Mathematics, University of Western Ontario, London, ON N6A 5B7, Canada\\
and \\
Department of Mathematics and
Computer Science, Algoma University, Sault St.Marie, ON P6A
2G4, Canada}

\maketitle

\maketitle

\begin{abstract}
We employ renormalization group (RG) summation techniques to obtain portions of Laplace QCD sum rules for scalar gluon currents beyond the order to which they have been explicitly calculated.  The first two of these sum rules are considered in some detail, and it is shown that they have significantly less dependence on the renormalization scale parameter $\mu^2$ once the RG summation is used to extend the perturbative results. Using the sum rules, we  then compute the bound on the scalar glueball mass and demonstrate that the 3 and 4-Loop perturbative results form lower and upper bounds to their RG summed counterparts. We further demonstrate improved convergence of the RG summed expressions with respect to perturbative results. 
\end{abstract}

\section*{Introduction} 
When computing the radiative corrections to physical processes, it is necessary to introduce a scale parameter $\mu^2$ in order to remove divergences through the renormalization procedure.  Exploiting the fact that any explicit dependence on $\mu^2$ must be cancelled by implicit dependence of physical parameters on $\mu^2$, one obtains the renormalization group (RG) equation.  This equation has been used to fix portions of radiative effects beyond those determined by direct perturbative calculation.  (For example, see refs. [1,2,3,4,5,6].)  In this letter we apply this approach to Laplace QCD sum rules for scalar gluonic currents.  We find that within RG summation, the dependence on the scale parameter $\mu^2$ is significantly diminished; this is expected as any exact solution of the RG equation would necessarily have no dependence on $\mu^2$. To illlustrate the physical significance of our approach, we do a full QCD sum rule calculation and apply it to determine the mass bound for the scalar glueball. The RG summation approach provided better results with improved convergence and lesser scheme dependence than those obtained via using purely perturbative inputs. This indicates that the RG summation approach can potentially be beneficial to other QCD sum rule applications as well.  

\section*{The Perturbative Laplace QCD Sum Rules}
The scalar gluonic correlation function is expressed as 

\begin{equation}
\Pi_{G} (p^2) = i\int d^4y\, e^{ip \cdot y} <0|T j_G(y) j_G(0)|0>
\end{equation}
where $j_G(y)=\frac{\beta(x)}{\alpha_s\beta_0} G{^{a}}_{\mu\nu}(y)G^{a,\mu\nu}(y)$, $x = \alpha_s/\pi$ and $\beta(x)$ is the QCD $\beta$-function defined for the evolution of the QCD strong coupling constant, $\alpha_s$.

 The perturbative Laplace sum rule $\mathcal{L}_k^{\mathrm{pert}}$ [7, 8, 9, 10] is given by
\begin{equation}
\mathcal{L}_k^{\mathrm{pert}} (\tau) = \frac{1}{\pi}\int_0^\infty ds\,s^{k+2} e^{-s\,\tau} Im\, \Pi_{G}^{\mathrm{pert}}(s) 
\end{equation}

where $\tau$ is the inverse square of the Borel mass. 
The imaginary part of the perturbative scalar gluonic correlator at centre of mass energy $s$, can be extracted from $Im <(G^2)^2>$, which has been computed to $\mathcal{O}(\alpha_s^4)$ in [11] and $\mathcal{O}(\alpha_s^5)$ in the QCD coupling $\alpha_s$ [12]. 
One can extract $Im \,\Pi^{\mathrm{pert}}_{G}(s)$ in the following way using the expression [13]
\begin{equation}
Im\; \Pi_G^{\mathrm{pert}}(s)=\frac{x^2}{\pi^2\beta_0^2}\left(\beta_0 + \beta_1 x + \beta_2 x^2 + \beta_3 x^3 \ldots \right)^2\,Im <(G^2)^2>= \frac{2 s^2x^2}{\pi^3} \left[ 1 + \sum_{n=1}^\infty \sum_{m=0}^n T_{n,m} x^n L^m(s) \right]
\end{equation}
where $L(s) = \log(s/\mu^2)$.
In ref. [13] the results to order $\mathcal{O}(\alpha_s^4)$ appear; here we make use of the following results to order $\mathcal{O}(\alpha_s^5)$ with 3 active quark flavours. 

\begin{align}
T_{1,0} &= \frac{659}{36} & T_{2,0} &= 197.515 & T_{3,0} &= 1349.88\\
         &                &  T_{2,1} &= -2105/16 & T_{3,1} &= -2107.42\notag\\
T_{1,1} &= -\frac{9}{2}   &                     &                    \notag \\
          &                & T_{2,2} &= 243/16 &   T_{3,2} &= 619.09 \notag \\
                          &                  &    T_{3,3} &= -45.56\;.\notag
\end{align} 
Together eqs. (2,3) lead to consideration of integrals of the form
\begin{equation}
J_m^{(k)} (a) = \int_0^\infty ds\,s^{k+2} e^{-s} \log^m(as)
\end{equation}

which satisfy
\begin{equation}
\frac{d}{da} J_m^{(k)}(a) = \frac{m}{a} J_{m-1}^{(k)} (a) .
\end{equation}
In particular we find that
\begin{subequations}
\begin{align}
J_0^{(0)} \left(\frac{1}{\tau\mu^2}\right) &= 2\\
J_1^{(0)} \left(\frac{1}{\tau\mu^2}\right) &= 3 - 2\gamma_E - 2 \log(\tau \mu^2)\\
J_2^{(0)} \left(\frac{1}{\tau\mu^2}\right) &= 2 + \frac{\pi^2}{3} - 6 \gamma_E +
2\gamma^2_E - 6 \log(\tau \mu^2)\\
&+ 4\gamma_E \log(\tau \mu^2) + 2\log^2 (\tau\mu^2)\nonumber\\
J_3^{(0)} \left(\frac{1}{\tau\mu^2}\right) &=  \frac{3\pi^2}{2} - \pi^2\gamma_E + 9\gamma_E^3 - 2\gamma_E^2 - 4\zeta(3) - 6\gamma_E\\
&+  \log(\tau \mu^2) \left[ 18 \gamma_E^2 - 6\gamma_E - \pi^2 - 6\right]\nonumber \\
&+  \log^2(\tau \mu^2) \left[ 9-6\gamma_E \right] - 2  \log^3(\tau \mu^2)\nonumber
\end{align}
\end{subequations}
where $\gamma_E$ = Euler's constant $= .5771 \ldots$.

Together, eqs. (4,7a-d) result in
\begin{align}
\mathcal{L}_0^{\mathrm{pert}} = \frac{4x^3}{\tau^3} &\Big[ 1 + x \left(T_{1,0}^{(0)} + T_{1,1}^{(0)} L^\prime\right) + x^2 \left( T_{2,0}^{(0)} + T_{2,1}^{(0)} L^\prime + T_{2,2}^{(0)} L^{\prime 2}\right)\nonumber \\
& + x^3\left(T_{3,0}^{(0)} + T_{3,1}^{(0)} L^\prime + T_{3,2}^{(0)} L^{\prime 2} + T_{3,3}^{(0)} L^{\prime 3}\right)\nonumber \\
& + \ldots \Big] \quad (L^\prime \equiv \log (\tau \mu^2))
\end{align}
where
\begin{align}
T_{1,0}^{(0)} &= 14.153  & T_{2,1}^{(0)} &= 103.53  & T_{3,1}^{(0)} &= 1135.32\\
T_{1,1}^{(0)} &= 4.5   & T_{2,2}^{(0)} &= 15.1875  & T_{3,2}^{(0)} &= 492.90\notag\\
T_{2,0}^{(0)} &= 95.042  & T_{3,0}^{(0)} &= 98.195  &   T_{3,3}^{(0)} &= 45.56.\notag
\end{align}
In a similar fashion we find that
\begin{subequations}
\begin{align}
J_0^{(1)} \left( \frac{1}{\tau\mu^2} \right) &= 6 \\
J_1^{(1)} \left( \frac{1}{\tau\mu^2} \right) & = 11 - 6\gamma_E - 6 \log (\tau\mu^2)\\
J_2^{(1)} \left( \frac{1}{\tau\mu^2} \right) & = 12 - 22\gamma_E + 6\gamma_E^2 + \pi^2 - 2 (11- 6\gamma_E) \log(\tau\mu^2) \\
& \qquad + 6\log^2(\tau\mu^2)\nonumber \\
J_3^{(1)} \left( \frac{1}{\tau\mu^2} \right) &= 6-12 \zeta (3) - 6\gamma_E^3 - 36\gamma_E + \frac{11}{2} \pi^2 + 33\gamma_E^2 - 3\pi^2\gamma_E \nonumber \\
&\qquad -3 (12-22\gamma_E + 6\gamma_E^2 + \pi^2) \log (\tau\mu^2) \\
& \qquad +3 (11-6\gamma_E) \log^2 (\tau \mu^2) -6\log^3(\tau\mu^2).\nonumber
\end{align}
\end{subequations}
Eqs. (4,10a-d) together lead to
\begin{align}
\mathcal{L}_1^{\mathrm{pert}}& = \frac{12x^2}{\tau^4}\Big[ 1 + x\left(T_{1,0}^{(1)} + T_{1,1}^{(1)} L^\prime\right) + x^2 \big( T_{2,0}^{(1)}+ T_{2,1}^{(1)} L^\prime \nonumber \\
& \qquad + T_{2,2}^{(1)} L^{\prime 2} \big) + x^3 \left( T_{3,0}^{(1)}+ T_{3,1}^{(1)} L^\prime + T_{3,2}^{(1)} L^{\prime 2} + T_{3,3}^{(1)} L^{\prime 3}\right) + \ldots \Big]
\end{align}
where
\begin{align}
T_{1,0}^{(1)} &= 12.653 \quad & T_{2,1}^{(1)} &= 93.4079 \quad & T_{3,1}^{(1)}& = 806.7219\\
T_{1,1}^{(1)} &= 4.5 \quad   & T_{2,2}^{(1)} &= \frac{243}{16} \quad & T_{3,2}^{(1)} &= 447.3438\notag\\
T_{2,0}^{(1)} &= 60.5312 \quad & T_{3,0}^{(1)} &= -280.2466 \quad &   T_{3,3}^{(1)} &= 45.56.\notag
\end{align}

\section*{The RG Summed Laplace QCD Sum Rules}
We now define
\begin{align}
S^{(k)} = 1 + \sum_{n=1}^\infty \sum_{m=0}^n \quad T_{n,m}^{(k)} x^n \log^m & (\tau \mu^2)\\
& (i = 1,2 \ldots)\nonumber
\end{align}
so that
\begin{equation}
\mathcal{L}_k^{\mathrm{pert}} = A_k \frac{x^2}{\tau^{k+3}} S^{(k)}
\end{equation}
by eq. (2).  So also by eq. (2)
\begin{equation}
\frac{d}{d\tau} \mathcal{L}_k^{\mathrm{pert}} = - \mathcal{L}_{k+1}^{\mathrm{pert}}
\end{equation}
and so by eqs. (13-15)
\begin{subequations}
\begin{align}
A_{k+1} &= (k + 3)A_k \\
(k+3)T_{n,m}^{(k+1)} &= (k+3) T_{n,m}^{(k)} - (m+1) T_{n,m+1}^{(k)}
\end{align}
\end{subequations}
showing that $S^{(k+1)}$ is fixed by  $S^{(k)}$.

Regrouping terms in the sum in eq. (13), we can write
\begin{equation}
S^{(k)} = \sum_{n=0}^\infty x^n S_n^{(k)} (U)
\end{equation}
where
\begin{equation}
S_n^{(k)} (U) = \sum_{m=0}^\infty T_{n+m,m}^{(k)}\, U^m \quad \quad (T_{00}^{(k)} = 1)
\end{equation}
where $U \equiv x \log (\tau\mu^2) = x L$.  $S_0^{(k)}$ is the leading-log (LL) contribution to $\mathcal{L}_k^{\mathrm{pert}}$, $S_1^{(k)}$ the next-to-leading-log (NLL) contribution $\ldots$ $S_{p}^{(k)}$ the $N^pLL$ contribution.

Since the explicit and implicit dependence of $\mathcal{L}_k^{\mathrm{pert}}$ on the unphysical parameter must cancel, we have the RG equation
\begin{equation}
\mu^2 \frac{d}{d\mu^2} \mathcal{L}_k^{\mathrm{pert}} = 0
\end{equation}
which by eq. (14) becomes
\begin{equation}
\left[ \beta (x) \left( \frac{2}{x} + \frac{\partial}{\partial x}\right) + \frac{\partial}{\partial L}\right] S^{(k)} = 0
\end{equation}
where we have the QCD $\beta$-function
\begin{equation}
\mu^2 \partial x /\partial\mu^2 = \beta(x) = -x^2 \left(\beta_0 + \beta_1 x + \beta_2 x^2 + \beta_3 x^4 \ldots \right)
\end{equation}
where $\beta_0=9/4$, $\beta_1=4$, $\beta_2=10.06$ and $\beta_3=47.23$ for 3 active flavours. 
(Note: The anomalous dimension $\gamma=0$ for scalar gluonic currents.)\\
Order-by-order in powers of $x$, eqs. (17,20) lead to
\begin{subequations}
\begin{align}
(1 - \beta_0 U) S_0^{\prime^{(k)}} &- 2\beta_0 S_0^{(k)} = 0\\
(1 - \beta_0 U) S_1^{\prime^{(k)}} - &3 \beta_0  S_1^{(k)} - \beta_1 \left(2 + U \frac{d}{dU}\right) S_0^{(k)} = 0\\
(1 - \beta_0 U) S_2^{\prime^{(k)}} - 4 \beta_0  S_2^{(k)} - & \beta_1\left(3 + U \frac{d}{dU}\right) S_1^{(k)} - \beta_2 \left(2 + U \frac{d}{dU}\right)  S_0^{(k)} = 0\\
\mathrm{etc.}\nonumber
\end{align}
\end{subequations}
The boundary conditions for these nested equations are
\begin{equation}
S_n^{(k)} (U = 0) = T_{n,0}^{(k)} .\qquad (n = 0, 1, \ldots)
\end{equation}
Solving eqs. (22a-c) in turn, we obtain
\begin{subequations}
\begin{align}
S_0^{(k)} = \frac{1}{w^2}& \quad (w = 1 - \beta_0 U)\\
S_1^{(k)} = \frac{1}{w^3} &\left( T_{1,0}^{(k)} - \frac{2\beta_1}{\beta_0}\ln \mid w \mid\right)\\
S_2^{(k)} =  \frac{1}{w^4} \big[  T_{2,0}^{(k)}& - \frac{\beta_1}{\beta_0} \left( 3 T_{1,0}^{(k)} + \frac{2\beta_1}{\beta_0 }\right) \ln \mid w \mid \nonumber\\
& \quad + \left( \frac{2\beta_1^2}{\beta^2 } - \frac{2\beta_2}{\beta_0 }\right) (w-1) + \frac{6\beta_1^2}{\beta_0} \ln^2 \mid w \mid \big].
\end{align}
\end{subequations}
(For $S_3^k$, see the appendix.)\\
To obtain $S_n^{(k)}(n >3)$ exactly, one needs $T_{n,0}^{(k)} (n > 3)$ and $\beta_n(n>3)$, neither of which have been computed as this involves five loop calculations.  However one could in the approximation $T_{n,0}^{(k)} = \beta_n = 0 (n > 3)$ solve for $S_n^{(k)} (n > 3)$.

Using explicit numerical values of the parameters occurring in eqs. (24, A.3) we find that with three quark flavours
\begin{equation}
w = 1 - \frac{9}{4} x \log (\tau\mu^2)
\end{equation}
\begin{subequations}
\begin{align}
S_0^{(k)}& = \frac{1}{w^2} \\
S_1^{(k)}& = \frac{1}{w^3} \left( T_{1,0}^{(k)} - \frac{32}{9} \ln \mid w \mid \right)\\
S_2^{(k)}& = \frac{1}{w^4} \big( T_{2,0}^{(k)} - 2.62115 (w-1) - \frac{512}{81} \ln \mid w \mid \\
   & \qquad - \frac{16}{3} T_{1,0}^{(k)} \ln \mid w \mid + \frac{256}{27} \ln^2 \mid w \mid\ \big) \nonumber \\
S_3^{(k)}& = \frac{1}{w^5} \big[ 38.964 + 24.9219\; T_{1,0}^{(k)} - 4.65981 w + T_{3,0}^{(k)}\\
& - 3.93172w \;T_{1,0}^{(k)} - 48.0277w^2 - 20.9902 \;T_{1,0}^{(k)} w^2 + 13.9935w^2\nonumber \\
& - 28.8766 \ln\mid w \mid - 9.4818\;  T_{1,0}^{(k)} \ln \mid w \mid - 7.11111\; T_{2,0}^{(k)} \ln \mid w \mid \nonumber \\
& + 13.9794 w \ln\mid w \mid + 74.6319 w^2 \ln \mid w \mid + 39.3306 \ln^2 \mid w \mid \nonumber \\
& + 18.963\; T_{1,0}^{(k)} \ln^2 \mid w \mid - 22.4746 \ln^3 \mid w \mid \big].\nonumber
\end{align}
\end{subequations}

An explicit four loop calculation with three quark flavours and taking $\Lambda_{QCD} = 300 MeV $leads to [14]
\begin{align}
\alpha_s(\mu^2)  = \frac{1}{\beta_0t} & \Big( 1 - \frac{\beta_1}{\beta_0} \frac{\ln t}{t} + \frac{\beta_1^2(\ln^2 t- \ln t - 1)+\beta_0\beta_2}{\beta_0^4 t^2}\\
& - \frac{\beta_1^3 (\ln^3t - \frac{5}{2} \ln^2t - 2 \ln t + \frac{1}{2}) + 3\beta_0\beta_1\beta_2 \ln t - \frac{1}{2} \beta_0^2 \beta_3}{\beta_0^6 t^3}\Big)\nonumber
\end{align}
where $t = \ln(\mu^2/\Lambda^2_{QCD})$.\\
We plot the purely perturbative $\mathcal{L}_0^{\mathrm{pert}}$ and $\mathcal{L}_1^{\mathrm{pert}}$ of eqs. (8, 11) with the RG improved expressions following from eqs. (14,17,26,27) in figs. 1, 2, 3 and 4. For parametrizing $\mu$ dependence, we define $\mu=\frac{\xi}{\sqrt{\tau}}$, and plot perturbative and RG-summed expressions for phenomenologically relevant values of $\xi=$ 0.8, 1 and 1.2 respectively.  In both sum rules, we note that the RG summed values are remarkably less renormalization scale dependent than the fixed order perturbative results.

To demonstrate the usefulness of our approach, we compute the mass of the scalar glueball using both purely perturbative and RG summation results. Utilizing a standard QCD sum rule approach as in ref. [10], we incorprate non-perturbative parts which include condensate and instanton contributions to the Laplace sum rules. These pieces are combined as follows,

\begin{equation}
\mathcal{L}_k=\mathcal{L}_k^{\mathrm{pert/RG}}+\mathcal{L}_k^{\mathrm{cond}}+\mathcal{L}_k^{\mathrm{inst}},
\end{equation}
where $k=0,1$ and the second and third term are condensate and instanton contributions respectively. We use the provided expressions for $\mathcal{L}_k^{\mathrm{cond}}$ and $\mathcal{L}_k^{\mathrm{cond}}$ in [10] and use the same set of QCD input parameters. The sum-rules provide a robust upper bound on the scalar glueball mass $m$ 
\begin{equation}
m\le\sqrt{\frac{\mathcal{L}_1}{\mathcal{L}_0}}\,~.
\end{equation}

In Figure 5, we plot the mass bound computed from both perturbative and RG summed Laplace sum rules. We not only find reduced scale dependence for the RG summed expressions, but also note that the purely 3-Loop and 4-Loop estimates are upper bounds to the RG summed mass estimates. This amply demonstrates (using a full QCD sum rule calculation) the benefit of using RG-summed expressions, as compared to using the purely perturbative results.\\

Towards demonstrating the convergence properties, we plot the 3-loop and 4-loop mass estimates separately, both for perturbative and RG-summed results. Figures 6 and 7 indicate better convergence properties of the RG summed results.  

Finally, we also propose an alternate rearrangement of the sum in eq. (13), so that in place of eq. (17) we have
\begin{equation}
S^{(k)} = \sum_{m=0}^\infty a_m^{(k)} (x) L^m,
\end{equation}
where
\begin{equation}
a_m^{(k)} = \sum_{n=0}^\infty T_{m+n,m}^{(k)} x^{n+m} .
\end{equation}
Substitution of eq. (30) into eq. (20) shows that the RG equation is satisfied at each order in $L$ provided
\begin{equation}
a_{n+1}^{(k)} = - \frac{\beta(x)}{n+1} \left( \frac{2}{x} + \frac{d}{dx}\right) a_n^{(k)} (x).\quad  (n = 0,1, \ldots)
\end{equation}
If now
\begin{equation}
a_n^{(k)} (x) = \left[ \exp\left( - 2 \int^x \frac{d\tilde{x}}{\tilde{x}}\right)\right] b_n^{(k)} (x)
\end{equation}
and
\begin{equation}
\frac{dx}{d\eta} = \beta (x)
\end{equation}
then
\begin{equation}
b_n^{(k)} (\eta) = -\frac{1}{n} \frac{d}{d\eta} b_{n-1}^{(k)} (\eta) = \frac{(-1)^n}{n!} \left( \frac{d}{d\eta}\right)^n b_0^{(k)} (\eta).
\end{equation}
Together, eqs. (30-35) show that
\begin{align}
S^{(k)} &= \left[ \sum_{n=0}^\infty \frac{(-L)^n}{n!} \frac{d^n}{d\eta^n} b_0^{(k)}(\eta) \right] \exp  \left( -2 \int^x \frac{d\tilde{x}}{\tilde{x}}\right)\nonumber \\
& = a_0^{(k)} x (\eta - L).
\end{align}
(Changes in the boundary condition of eq. (34) can be compensated by changes in $\mu^2$ in $L$.)  Eq. (36) is not unexpected; it shows how all log-dependent contributions to $S^{(k)}$ are fixed by the RG equation to be given in terms of the log-independent contribution to $S^{(k)}$ (i.e., $a_0$).

\section*{Discussion}
Using the four loop $\beta$-function in QCD as well as the four loop contribution to the scalar gluonic correlation function, we have explicitly summed the $LL \ldots N^3LL$ contribution to the corresponding Laplace QCD sum rules.  By having incorporated these contributions, the sum rules $\mathcal{L}_0^{\mathrm{pert}}$ and $\mathcal{L}_1^{\mathrm{pert}}$ have a considerably reduced dependence on the non-physical renormalization scale $\mu^2$.

It is also possible to use the RG equation to show how all log-dependent contributions to the Laplace sum rules are fixed by the log-independent contributions.

\section*{Acknowledgements}
FC would like to thank the support and generous hospitality of the CERN Theory Group, where work on this paper was conducted. TGS would like to thank Natural Science and Engineering Research Council (NSERC) for their support. R. McLeod provided useful discussions. 

\section*{Appendix}
The equation for $S_3^{(k)}$ that follows from eq. (17,20,21) is
\[
\left[ (1 - \beta_0 U) \frac{d}{dU} -  5 \beta_0\right] S_3^{(k)} - \beta_1 \left[ 4 + U \frac{d}{dU}\right] S_2^{(k)} - \beta_2 \left[ 3 + U \frac{d}{dU}\right] S_1^{(k)} \nonumber \]
\[ \qquad - \beta_3 \left[ 2 + U \frac{d}{dU}\right] S_0^{(k)}= 0. \eqno(A.1) \]
Writing eqs. (24a,b) as
\[
S_1^{(k)} = \frac{1}{w^3} (A + B \ln \mid w \mid), \quad
S_2^{(k)} = \frac{1}{w^4} (C + D(w+1) + E \ln \mid w \mid + F \ln^2 \mid w \mid)
\eqno(A.2a,b)\]
it is easily shown that the solution to eq. (A.1) is
\[
S_3^{(k)} = \frac{1}{w^5} \big[ T_{3,0}^{(k)} - \left( \frac{\beta_1}{\beta_0} (4C - 4D - E)\right) \ln \mid w\mid - \left( \frac{\beta_1}{\beta_0} (3D + E) + \frac{\beta_2}{\beta_0} (3A -B)\right) (w-1)\nonumber \]
\[ - \left( \frac{\beta_1}{\beta_0} D + \frac{\beta_2}{\beta_0} B + \frac{2\beta_3}{\beta_0}\right) (w-1) - \left( \frac{\beta_1}{\beta_0} (4E - 2F)\right) \frac{\ln^2\mid w \mid}{2} \eqno(A.3) \]
\[ \qquad - \left( \frac{\beta_1}{\beta_0} (2F) +  \frac{\beta_2}{\beta_0} (3B)\right) \left( w \ln \mid w \mid - (w-1) \right) - \left( \frac{\beta_1}{\beta_0} (4F) \right)
\frac{\ln^3\mid w \mid}{3} \big].\nonumber \]

\begin{figure}[hbt]
\begin{center}
\includegraphics[scale=0.9]{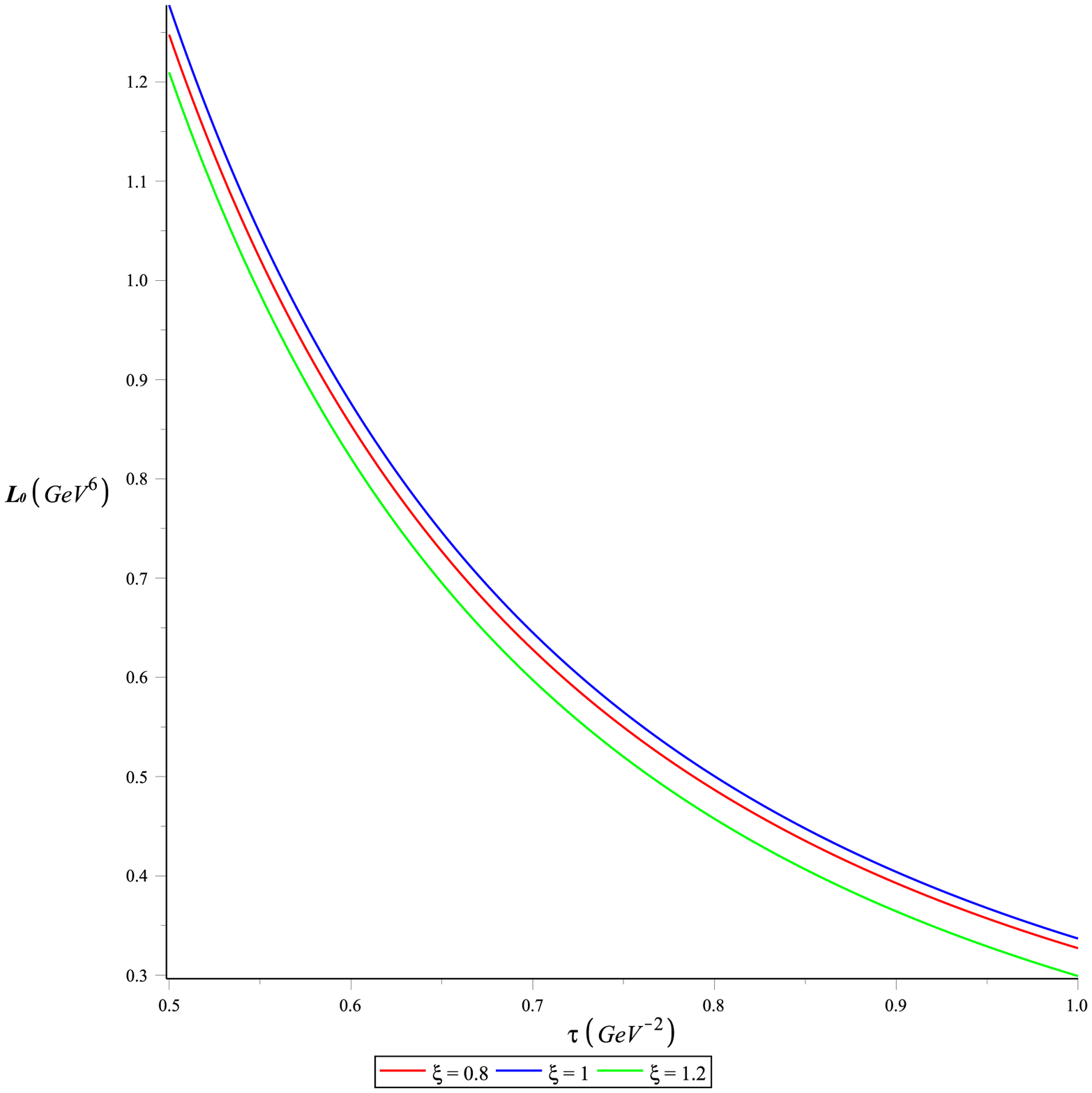}
\caption{The $\mu$ dependence of the purely perturbative sum rule $\mathcal{L}_0^{\mathrm{pert}}$  $(GeV^6)$ with respect to $\tau$ $(GeV^{-2})$using values of $\xi=$ 0.8, 1 and 1.2 respectively}
\label{Fig.1}
\end{center}
\end{figure}

\begin{figure}[hbt]
\begin{center}
\includegraphics[scale=0.9]{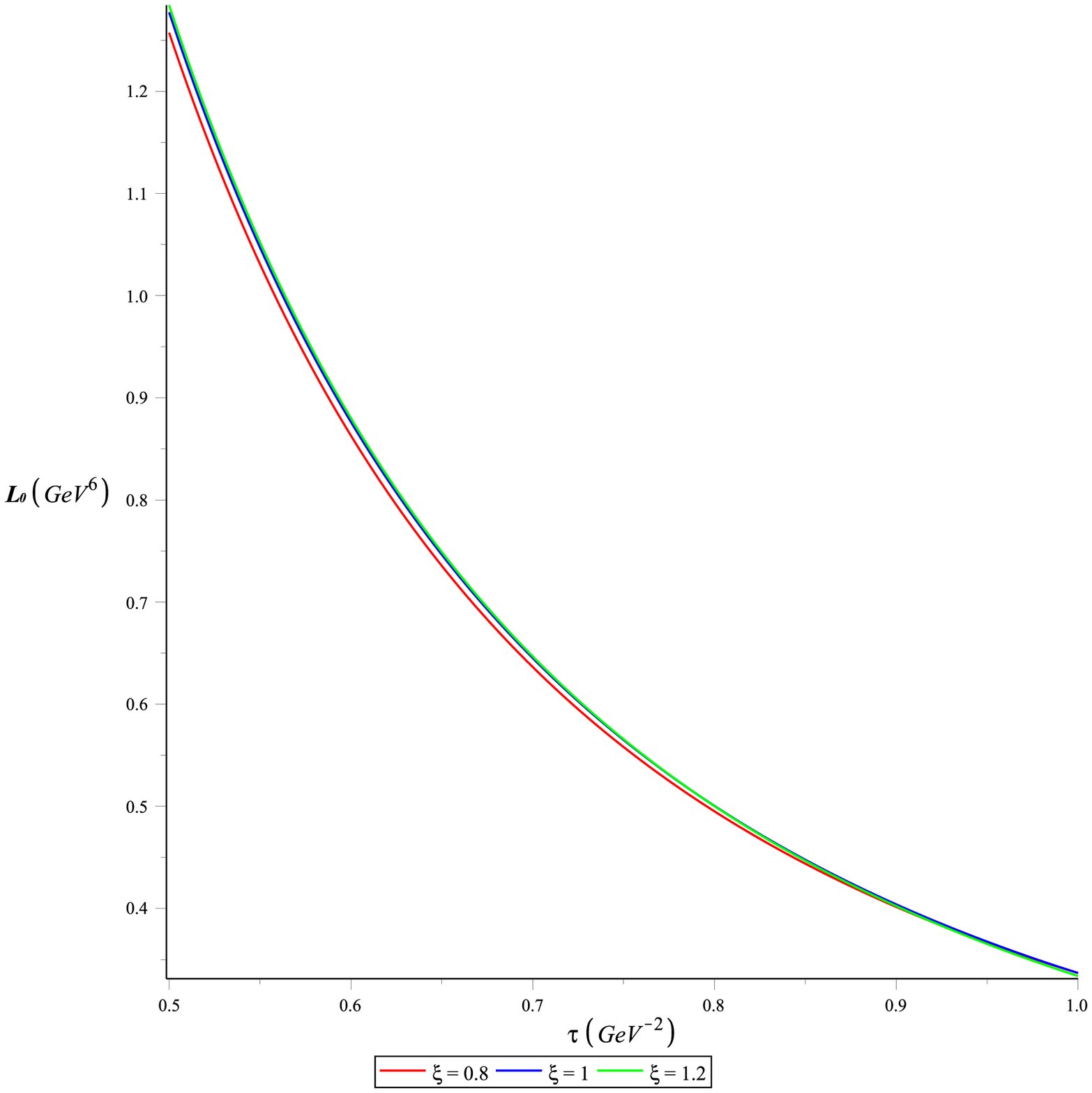}
\caption{The $\mu$ dependence of the RG-summed sum rule $\mathcal{L}_0^{\mathrm{pert}}$  $(GeV^{6})$ with respect to $\tau$ $(GeV^{-2})$ using values of $\xi=$ 0.8, 1 and 1.2 respectively}
\label{Fig. 2}
\end{center}
\end{figure}

\begin{figure}[hbt]
\begin{center}
\includegraphics[scale=0.9]{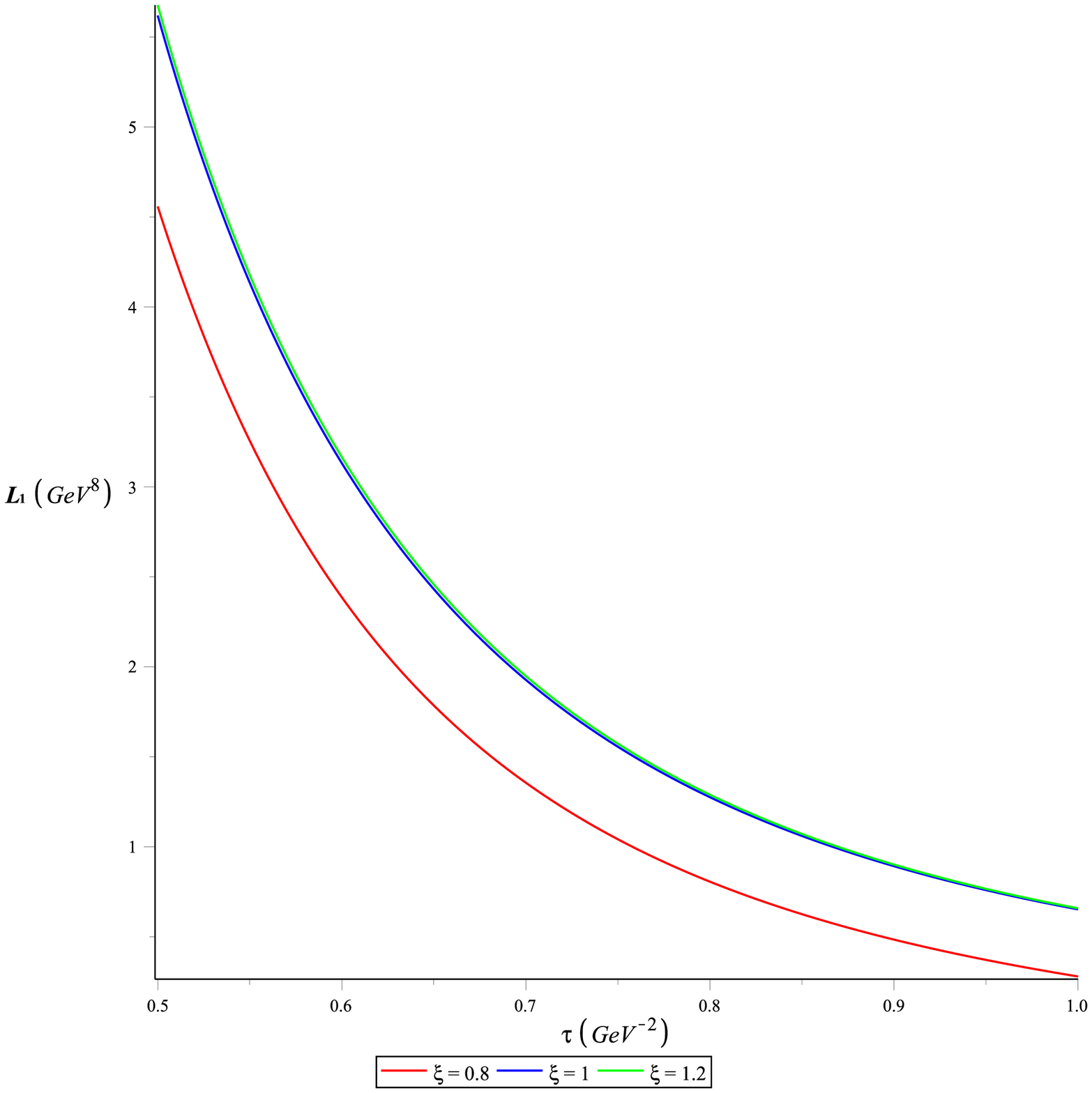}
\caption{The $\mu$ dependence of the purely perturbative sum rule $\mathcal{L}_1^{\mathrm{pert}}$  $(GeV^{8})$ with respect to $\tau$ $(GeV^{-2})$ using values of $\xi=$ 0.8, 1 and 1.2 respectively}
\label{Fig. 3}
\end{center}
\end{figure}

\begin{figure}[hbt]
\begin{center}
\includegraphics[scale=0.9]{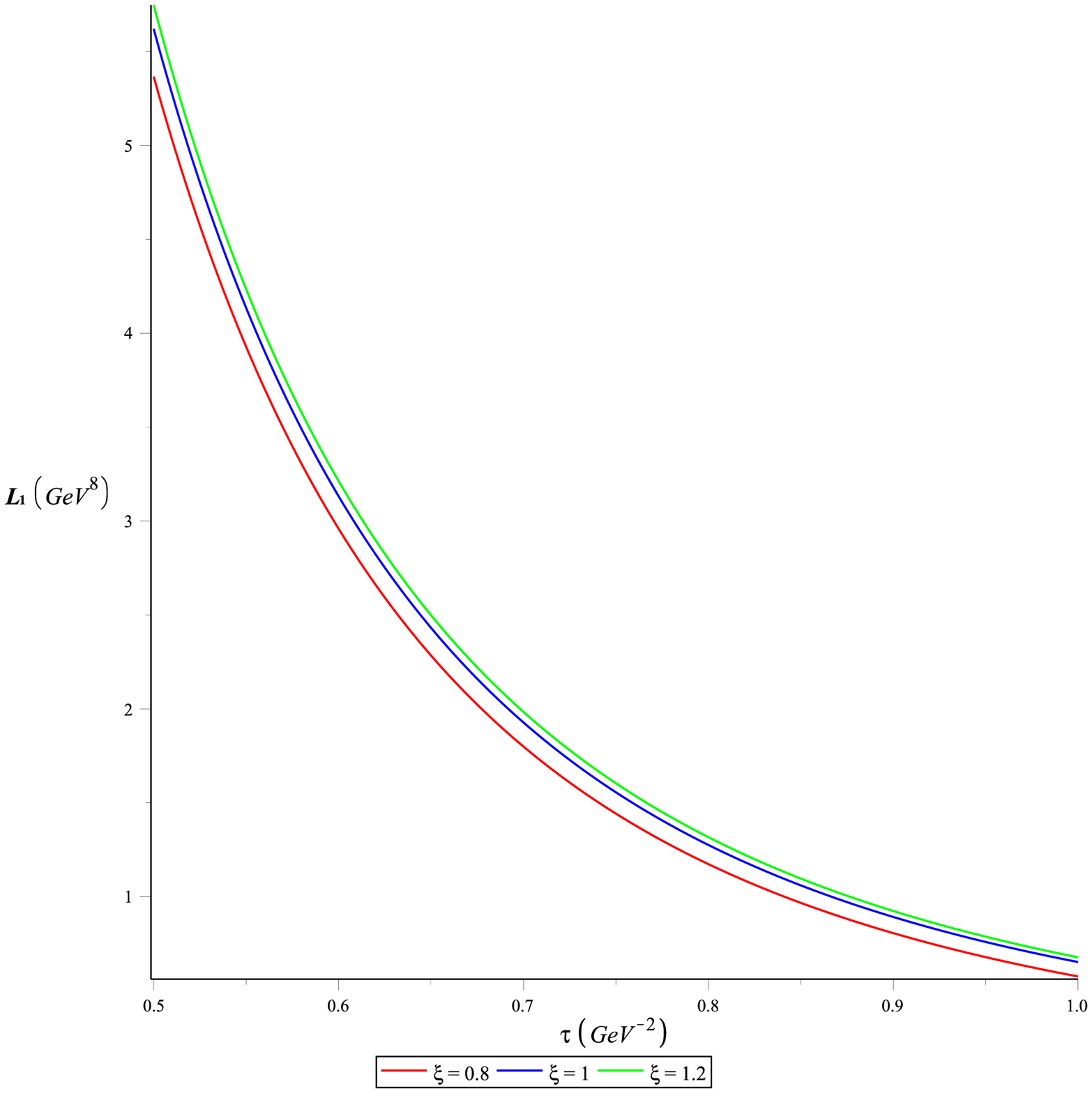}
\caption{The $\mu$ dependence of the RG-summed sum rule $\mathcal{L}_1^{\mathrm{pert}}$  $(GeV^{8})$ with respect to $\tau$  $(GeV^{-2})$ using values of $\xi=$ 0.8, 1 and 1.2 respectively}
\label{Fig. 4}
\end{center}
\end{figure}

\begin{figure}[hbt]
\begin{center}
\includegraphics[scale=1]{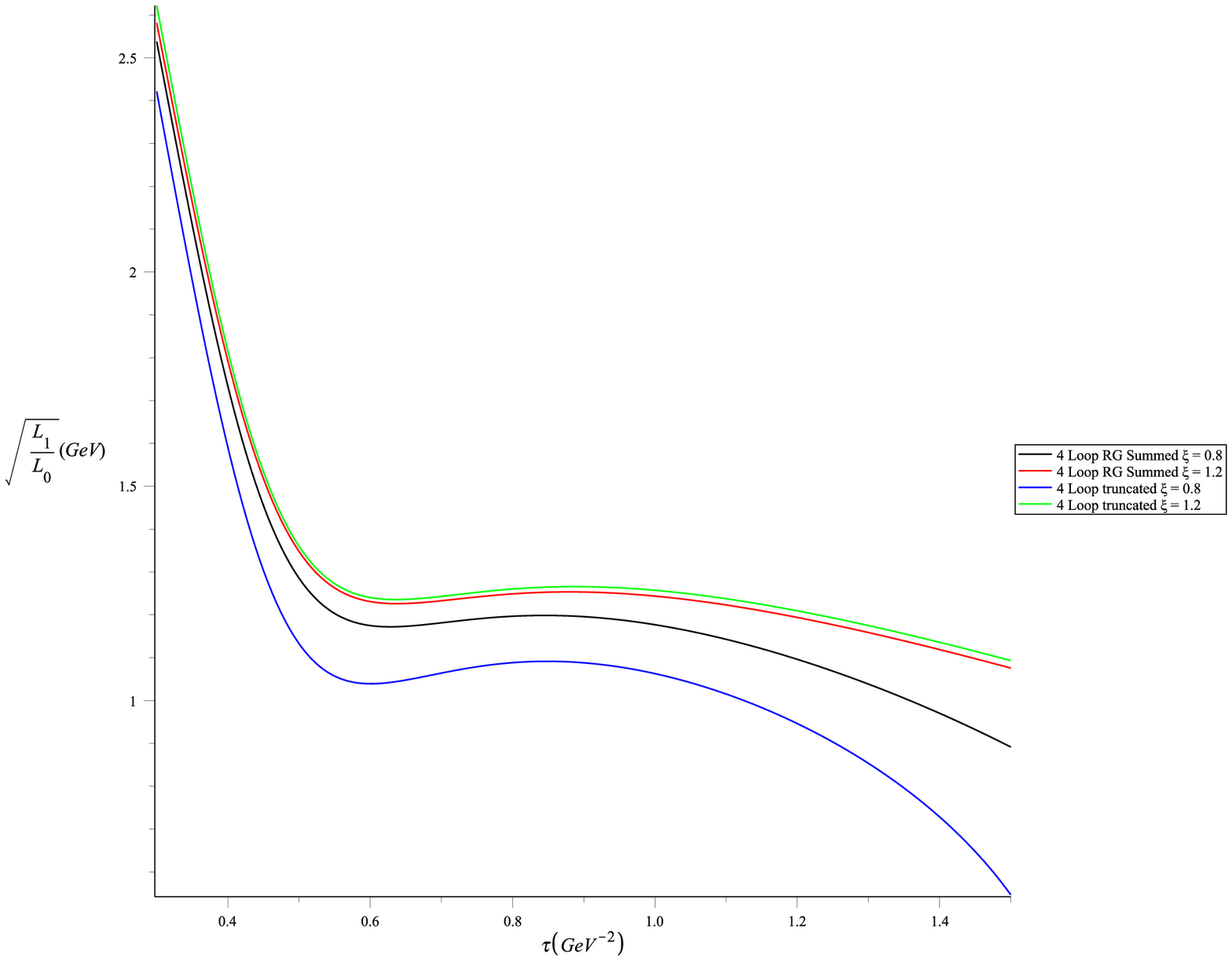}
\caption{The $\mu$ dependence of the scalar glueball mass bound in both truncated and RG summed form with respect to $\tau$  $(GeV^{-2})$ using values of $\xi=$ 0.8 and 1.2 respectively}
\label{Fig. 5}
\end{center}
\end{figure}

\begin{figure}[hbt]
\begin{center}
\includegraphics[scale=1]{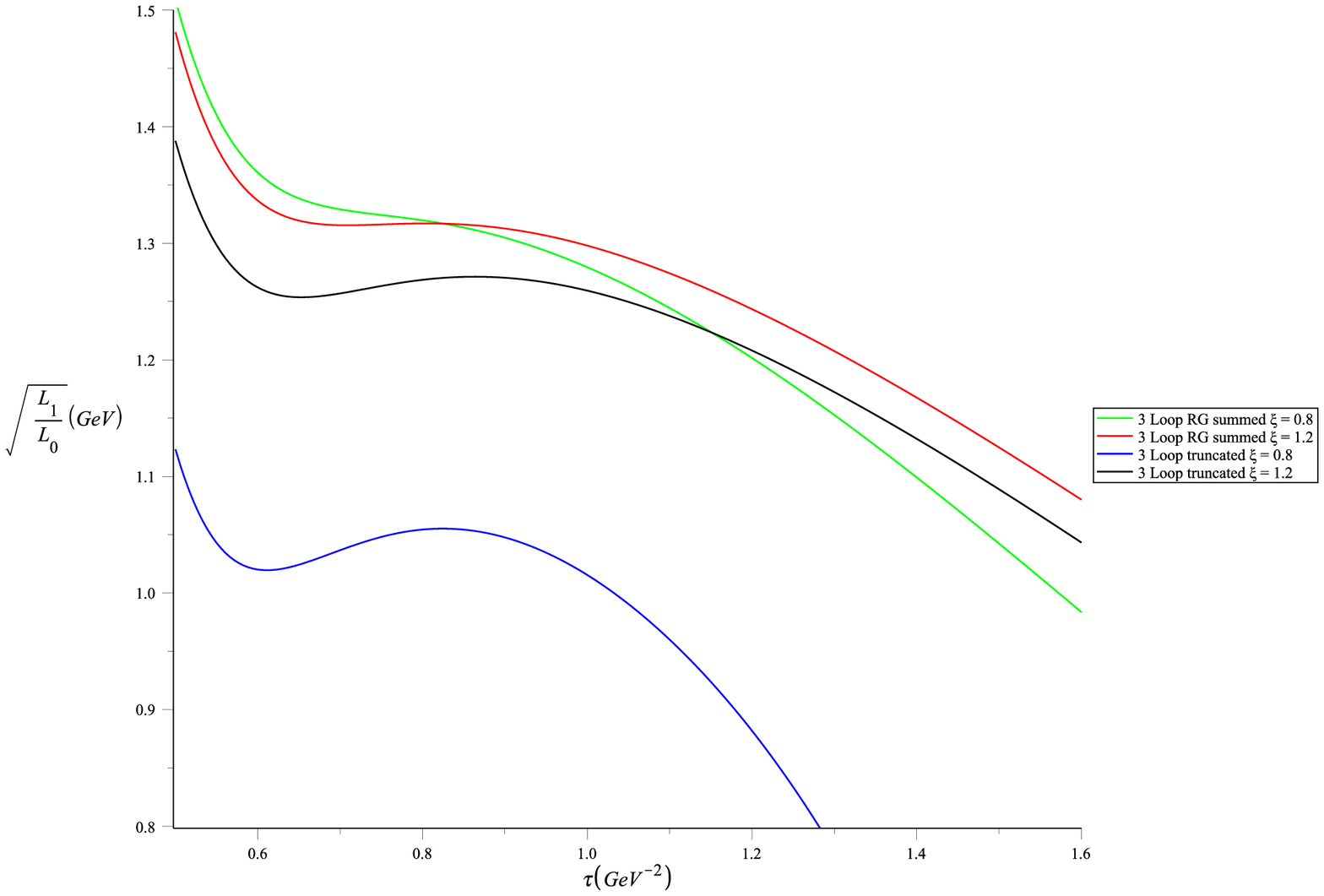}
\caption{The 3-Loop $\mu$ dependence of the scalar glueball mass bound in both truncated and RG summed form with respect to $\tau$  $(GeV^{-2})$ using values of $\xi=$ 0.8 and 1.2 respectively}
\label{Fig. 6}
\end{center}
\end{figure}

\begin{figure}[hbt]
\begin{center}
\includegraphics[scale=1]{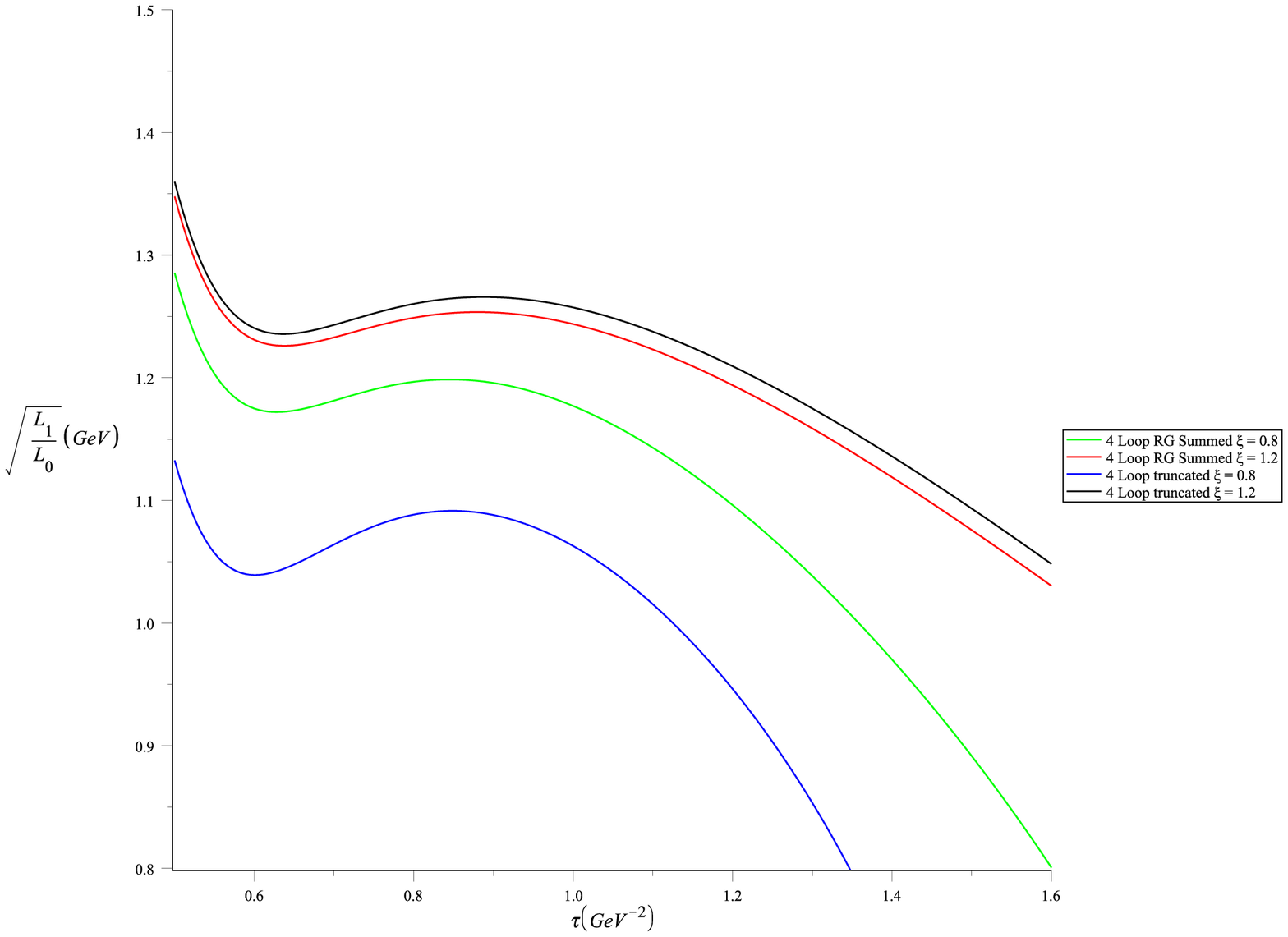}
\caption{The 4-Loop $\mu$ dependence of the scalar glueball mass bound in both truncated and RG summed form with respect to $\tau$  $(GeV^{-2})$ using values of $\xi=$ 0.8 and 1.2 respectively}
\label{Fig. 7}
\end{center}
\end{figure}

\end{document}